\documentclass[pra,aps,twocolumn,nofootinbib,floatfix,showpacs]{revtex4}

\newcommand{\ybg}{\textsuperscript{172}Yb}

\newcommand{\ybu}{\textsuperscript{171}Yb}
\newcommand{\ybuion}{\textsuperscript{171}Yb\textsuperscript{+}}

\usepackage{ulem}
\usepackage{textcomp}
\usepackage{graphicx}
\usepackage{color}

\usepackage{bm,amsmath}
\usepackage{bbm}
\usepackage{times,txfonts,hyperref}

\newcommand{\rp}[1]{(\ref{#1})}

\newcommand{\abs}[1]{\left|{#1}\right|}

\newcommand{\av}[1]{\left\langle #1 \right\rangle}

\newcommand{\br}[1]{\langle #1|}

\newcommand{\ke}[1]{|#1\rangle}

\newcommand{\al}[1]{^{(#1)}}
\newcommand{\da}{^\dagger}

\newcommand{\pt}[1]{\left( #1 \right)}
\newcommand{\pq}[1]{\left[ #1 \right]}
\newcommand{\pg}[1]{\left\{ #1 \right\}}

\newcommand{\rbar}[1]{\left. #1 \right|}

\newcommand{\ee}{{\rm e}}
\newcommand{\ii}{{\rm i}}

\newcommand{\nn}{{\nonumber}}

\newcommand{\LL}{{\cal L}}

\begin{document}

\title[Long distance entanglement in a segmented ion trap]{
Adiabatic quantum simulation with a segmented ion trap: Application to long-distance entanglement in quantum spin systems}

\author{S. Zippilli$^1$, M. Johanning$^2$, S. M. Giampaolo$^{1,3}$, Ch. Wunderlich$^2$,
and F. Illuminati$^{1,4,5}$\footnote{Corresponding authors: illuminati@sa.infn.it, christof.wunderlich@uni-siegen.de}}
\affiliation{$^{1}$\mbox{Dipartimento di Ingegneria Industriale, Universit\`a degli Studi di Salerno, Via Giovanni Paolo II 132, I-84084 Fisciano (SA), Italy}
\\
$^{2}$\mbox{Fachbereich Physik, Naturwissenschaftlich-Technische Fakult\"at, Universit\"at Siegen, 57068 Siegen, Germany}
\\
$^{3}$\mbox{University of Vienna, Faculty of Physics, Boltzmanngasse 5, 1090 Vienna, Austria}
\\
$^{4}$\mbox{INFN, Sezione di Napoli, Gruppo Collegato di Salerno, I-84084 Fisciano (SA), Italy}
\\
$^{5}$\mbox{CNISM Unit\`a di Salerno, I-84084 Fisciano (SA), Italy
}}

\date{January 31, 2014}


\begin{abstract}

We investigate theoretically systems of ions in segmented linear Paul traps for the quantum simulation of quantum spin models with tunable interactions. The scheme is entirely general and can be applied to the realization of arbitrary spin-spin interactions. As a specific application we discuss in detail the quantum simulation of models that exhibit long-distance entanglement in the ground state. We show how tailoring of the axial trapping potential allows for generating spin-spin coupling patterns that are suitable to create long-distance entanglement. We discuss how suitable sequences of microwave pulses can implement Trotter expansions and realize various kinds of effective spin-spin interactions. The corresponding Hamiltonians can be varied on adjustable time scales, thereby allowing the controlled adiabatic preparation of their ground states.

\end{abstract}

\pacs{03.67.Ac, 37.10.Ty, 37.10.Vz}

\maketitle


\section{Introduction}
\label{sec:Introduction}

Entanglement is a  central resource for quantum technological applications \cite{BlattWineland2008,NielsenChuang2000}. Great effort has been devoted to the generation and distribution of entanglement between non-directly interacting systems, which can be either nodes of a quantum internet or distant elements inside a quantum computer \cite{Kimble,Cirac,Bose2003,
Zippilli08,
Zippilli2013,Campos Venuti1,Campos Venuti2,GiampaoloLong1,GiampaoloLong2,Gualdi,Zippilli}.
Particularly intriguing in this context is the prediction that certain spin models are naturally endowed with peculiar entanglement properties in their ground state which could be profitable for quantum communication purposes, for example, between different spatial regions within a quantum processor. Specifically, the concept of long-distance entanglement (LDE) has been introduced and discussed in order to identify the occurrence of sizeable nonlocal quantum correlations between distant, non-directly interacting spins in quantum spin chains and networks~\cite{Campos Venuti1,Campos Venuti2,GiampaoloLong1,GiampaoloLong2,Gualdi,Zippilli,Marzoli}. This phenomenon emerges in models with non degenerate ground states, when the end spins (spins at the boundary of the system) interact weakly with their immediate neighbors, such that a strongly correlated bulk mediates effective interactions between the distant,
non directly interacting, end spins. In this work we discuss the feasibility of schemes for the experimental observation of this effect using trapped ions as quantum simulators of quantum spin models.

Trapped ions are highly versatile systems which have been proven to be effective in quantum technological applications. The simulation of quantum models of strongly interacting quantum matter using trapped ions holds promise for the investigation of those quantum dynamics that remain so far unexplored due their inescapable complexity~\cite{Johanning2009,Schneider2012}. Indeed, the natural many-body dynamics of trapped atoms is very rich and interesting by itself; on the other hand, in the present work we will be mainly concerned with the subtle and intriguing task of realizing models that are not directly provided by the natural, i.e. non-engineered, physics of trapped ions. Although spin interactions emerge quite naturally in ion chain systems, engineering and control of a desired complex Hamiltonian can be a challenging task with high pay-off. Spectacular proof of principle experimental demonstrations~\cite{Friedenauer2008,Kim2010,Islam2011,Lanyon2011} have shown the potential of trapped ion based quantum simulators. However, so far none of these experiments has explored the ground state of spin models that are expected to exhibit highly non-classical properties. Here, we propose to implement spin-Hamiltonians with trapped ions taking advantage of the following features: i) shaping of the trapping potentials in order to suppress the effect of long-range interactions, ii) well controlled adiabatic processes driving the system to the ground state, and iii)  implementation of Trotterization (Trotter expansion) in order to generate the relevant spin-spin interactions in all  needed directions and components.

In what follows we will explore the capabilities of trapped ion systems for the quantum simulation of specific spin models, and we will apply them to propose the experimental demonstration of LDE in quantum spin chains.
LDE is a global nonclassical effect which, on the other hand, can be monitored by the analysis of only two spins, namely the end spins of the chain. It is therefore a sufficiently simple, yet rich phenomenon which is ideal to be demonstrated using an ion trap quantum simulator. Differing from the previous experiments cited above~\cite{Kim2010,Islam2011,Lanyon2011} in which the spin coherent manipulation is realized with laser fields, here we focus on segmented ion traps in the presence of a magnetic gradient where the engineering of the quantum dynamics is realized by microwave fields.

The paper is organized as follows. In Sec.~\ref{system} we introduce the systems and discuss the basic features of the scheme that we plan to implement for the simulation of  long distance entanglement. In Sec.~\ref{Quantum simulation} we discuss how to tailor the spin-spin interactions  and we describe the scheme of pulses (Trotter expansion) for the simulation of spin Hamiltonians. In Sec.~\ref{AdiabPulse} we discuss the results for the adiabatic preparation of the ground state and discuss the experimental feasibility of the protocol. Finally, in Sec.~\ref{sec:Outlook} we draw  conclusions  and discuss possible outlooks.

\section{The system}\label{system}

Doppler cooled ions held in a segmented ion trap \cite{Schulz2008,Kaufmann2011} and exposed to a magnetic field gradient realize effective spin-1/2 models~\cite{Mintert2001,Wunderlich2002,McHugh2005,Chiaverini2008,Johanning2009,Khromova2012}.
The effective spin-spin interactions induced by the magnetic field are of Ising type  and can be adjusted by tailoring the  axial trapping potential.
In particular, if the ions are sufficiently cold, such that the ion motion can be neglected (the validity of this approximation is discussed in Sec.~\ref{motion}), the effective system of $N$ spins is described by the Ising Hamiltonian
\begin{eqnarray}\label{barHIsing}
 \overline H_{\rm Ising}\al{z}&=&\overline H_z+ H_{zz}
 \nn\\
\overline H_z&=& \frac{\hbar}{2}\sum_{j=1}^N \omega_j \sigma_j^z
\nn\\
 H_{zz}&=& -\frac{\hbar}{2}\sum_{i,j} J_{ij} \sigma_i^z\sigma_j^z,
\end{eqnarray}
where the resonance frequencies of the atomic spins $\omega_j$ depend on the external magnetic field $B(x_{0,j})$ at  the equilibrium position of the ion $x_{0,j}$ \cite{BreitRabi1931}.
The spin-spin couplings are in general long range and their magnitude depends on the trapping potential and on the spatial derivative of the spin resonance frequency that, in turn, is determined by the magnetic field gradient.
They are given by
\begin{equation}
J_{ij}  = \frac{\hbar}{2}
\rbar{ \frac{\partial \omega_i}{\partial x_i} }_{x_{0,i}}
\rbar{ \frac{\partial \omega_j}{\partial x_j} }_{x_{0,j}} (A^{-1})_{ij},
\end{equation}
where  $A$, whose elements are
\begin{equation}
A_{ij}=\rbar{\frac{\partial^2 V(x_1,\cdots x_N)}{\partial x_i\ \partial x_j}}_{x_\ell=x_{0,\ell},\ \forall \ell} \ ,
\end{equation}
is the Hessian matrix of the potential energy function $V(x_1,\cdots x_N)$ that confines the ions with $x_j$ indicating the position of ion $j$.
In addition, the magnetic gradient allows for addressing of individual spins with a microwave field that can, therefore, be used to manipulate the spin dynamics~\cite{Mintert2001,Wunderlich2002,Khromova2012} (see also Sec.~\ref{spindynamics}).

\subsection{General considerations}

Spin Hamiltonians with non-trivial ground state correlations (as in the case of LDE) are in general  characterized by non-commuting spin-spin interaction terms. This is not the case for the simple Ising Hamiltonian~\rp{barHIsing} in which only terms of the form $\sigma_j^z\sigma_k^z$ are present. Therefore, the simulation of LDE requires the ability to engineer interactions along a different axes, described for example by a term of the form $\sigma_j^x\sigma_k^x$.
Such an effective interaction can be induced  using a sequence of $\pi/2$ microwave pules
that realize the transformation $\ee^{-\ii\pi\sigma_j^y/4}\sigma_j^z\ee^{\ii\pi\sigma_j^y/4}=\sigma_j^x$ over all the spins \cite{Hodges2007}. In particular,
a free evolution sandwiched by two trains of $\pi/2$ pulses (each pulse addressing a particular ion $j=1, ..., N$) with opposite phases performs the following  transformation
\begin{eqnarray}
\ee^{-\ii\frac{\pi}{4}\sigma_N^y}\cdots\ee^{-\ii\frac{\pi}{4}\sigma_1^y}\ee^{-\ii\overline H_{\rm Ising}\al{z}t}\ee^{\ii\frac{\pi}{4}\sigma_1^y}\cdots\ee^{\ii\frac{\pi}{4}\sigma_N^y}= \ee^{-\ii\overline H_{\rm Ising}\al{x}t},
\end{eqnarray}
(where $\overline H_{Ising}\al{x}$ is equal to the Hamiltonian in Eq.~\rp{barHIsing}, with all the operators $\sigma_j^z$ replaced with the corresponding $\sigma_j^x$)
and  realizes an Ising interaction along the $x$-axes.
In order for this transformation to be effective, the duration of the pulses have to be sufficiently short so that the evolution due to the spin-spin interactions can be neglected during the pulse. This is achieved with a  sufficiently strong microwave driving field resulting in a Rabi frequency $\Omega\gg J_{ij}$. On the other hand each microwave pulse should operate on a single spin, and its effect on the other spins should be negligible. This imposes a limit on the maximum allowed intensity of the driving field $\Omega\ll\Delta$, where $\Delta$ indicates the frequency difference between neighboring spin resonances.

The simultaneous interaction along $z$ and $x$ can be simulated by Trotterization, namely by repeated, fast  application of the two kind of interactions~\cite{Lloyd}.
Provided that the interaction time $\tau/n$ is sufficiently small,  it is possible to approximate
 \begin{eqnarray}
\ee^{-\ii\pq{\overline H_{Ising}\al{x} +\overline H_{Ising}\al{z}} \tau}\simeq\pq{\ee^{-\ii\overline H_{Ising}\al{z}\tau/n}\ee^{-\ii\overline H_{Ising}\al{x}\tau/n}}^n,
\end{eqnarray}
and to generate a stroboscopic evolution which simulates a Hamiltonian that  is the sum of two Ising Hamiltonians with interactions along the two orthogonal axes.

We also note that typically the parameters in the Hamiltonian $ \overline H_{Ising}\al{z}$ defined in Eq.~\rp{barHIsing} are such that the spin-spin coupling strengths are much smaller then the single site energy, $J_{ij}\ll\omega_\ell$, which hence dominate the dynamics of this model. Nevertheless we note that we are interested in the situation in which the system is driven by a series of microwave pulses. In this case, as demonstrated in the next section, the relevant dynamics is that obtained in a reference frame rotating at the driving field frequency. In  this representation, the relevant single site energy is in fact given by the detuning $b=\omega_j-\overline \omega_j$ between spin resonance frequency ($\omega_j$) and driving field frequency ($\overline \omega_j$), which can therefore be adjusted and controlled during the dynamics.

These results can eventually be used for the adiabatic preparation of the ground state of, for example, XX Hamiltonians. The system is prepared initially in the ground state of a sufficiently simple Hamiltonian which is easy to prepare: In our case it consists of the ferromagnetic/fully polarized spin state that is the ground state of the Ising Hamiltonian with a finite magnetic field ($H_{initial}=\hbar\ b/2\sum_j\sigma_j^z-\hbar/2\sum_{i,j} J_{ij}\sigma_i^z\sigma_j^z$). Then, the effective magnetic field is slowly switched off ($b$ is reduced) while  the interaction along $x$ is turned on by tuning the relative duration of the evolutions under the two Hamiltonians $\overline H_{Ising}\al{x}$ and $\overline H_{Ising}\al{z}$. If the variation of the parameters is sufficiently slow, then the system remains in the instantaneous ground state.
And eventually it approaches the ground state of the final modified target Hamiltonian $H_{final}=-\hbar/2\sum_{i,j} J_{ij}\pt{\sigma_i^z\sigma_j^z+\sigma_i^x\sigma_j^x}$ where the effective magnetic field is zero and both interactions along $x$ and $z$ are present. This Hamiltonian exhibits ground state long distance entanglement when the end spins are weakly coupled to the bulk~\cite{Campos Venuti1,Campos Venuti2,GiampaoloLong1,GiampaoloLong2,Gualdi,Zippilli}.

However, in general the typical  harmonic trapping potential of linear ion traps induces long range interactions with maximum couplings at the end of the chain. Thus, in order to obtain ground state LDE, the trapping potential has to be carefully engineered and the end spins interactions have to be made weak. This can be realized with segmented micro-traps as discussed in Sec.~\ref{sec:TailoringCouplingConstantsInASegmentedTrap}.

\section{Engineering of spin Hamiltonians with trapped Ions}\label{Quantum simulation}

In this section we study how to manipulate the coupling strengths $J_{i,j}$  and how various kinds of spin Hamiltonians can be designed, which differ from the $\sigma_i^z \sigma_j^z$-interaction that arises naturally for strings of trapped ions exposed to a  magnetic gradient \cite{Mintert2001,Wunderlich2002}. To be specific, we present detailed calculations for an existing  micro-structured ion trap \cite{Schulz2008,Kaufmann2011}. The principles used to obtain the concrete results presented in what follows are, of course, applicable to other segmented traps with a magnetic gradient as well.

\subsection{Tailoring the coupling constants in a segmented trap}
\label{sec:TailoringCouplingConstantsInASegmentedTrap}

In what follows we will discuss how to generate the axial trapping potential which results in the coupling pattern desired for LDE. In Ref.  \cite{HWunderlich2009},too, coupling patterns were calculated for ions held in a micro-structured trap. However, in that treatment single ions (or ion chains) are located at the bottom of an approximately harmonic potential. Thus the separation of minima becomes large (on the order 200~\textmu m) and the coupling between different sites can become impractically small (Hz) for the purpose described in this article. In contrast, here we consider ions held in closely separated anharmonic wells, and we tune both the harmonic and anharmonic part to obtain the desired coupling pattern.

The effective potential can be written as

\begin{equation}
\phi_{\rm{eff}} =  \frac{P_{\rm{rf}}}{P_0} \phi_{\rm{rf}} + \sum_i \frac{U_{i}}{U_0} \phi_{i}
\label{eq:oneSegmentEquilibriumCondition}
\end{equation}

where $\phi_{\rm{rf}}$ is the effective  harmonic potential due to the presence of the rf trapping field at an rf power level $P_{\rm{rf}}=P_0$. $\phi_{i}$ is the dc potential originating from electrode $i$ set to the voltage $U_i=U_0$. The rf effective potential $\phi_{\rm{rf}}$ is almost zero on the axis of a linear segmented trap due to symmetry reasons and thus its influence on the axial potential is neglected in the following discussion.


With a given voltage pattern $\{U_i\}$ applied to the electrodes, and an initial guess of ion positions,
one can calculate equilibrium positions by minimizing the total energy. Note that several local minima of the total energy are possible, as the ions can be distributed differently over the wells of the potential. In addition, permutations of ion positions yield identical total energies.
After the equilibrium ion positions have been determined, we calculate the normal modes of an ion string, the Zeeman shifts of individual ions and the resulting coupling constants.


For small excursion $\Delta x_i$ of ion $i$ from the equilibrium position, the motion of the chain can be decomposed into normal modes, which is equivalent to say that the force $F_{ij}$ on ion $j$ depends linearly on the excursion pattern $\vec{\Delta x}$ as
\begin{equation}
\vec{F} = \hat{A}\vec{\Delta x}
\label{eq:linForce}
\end{equation}





\begin{figure}[ht]
\centering
\includegraphics[width=0.95\columnwidth]{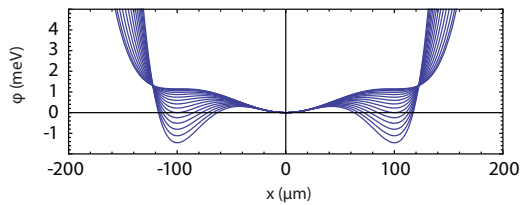}
\caption{Triple well potential as in Eq.~\ref{eq:tripleWell}, with $m$ being the mass of \ybuion, $x_o =$~1~\textmu m, $\omega_c = 2\pi \,\cdot$~100~kHz, and $\omega_o$ varied uniformly between 20 and 300~$\cdot\,2\pi$~kHz, where the curves with the most pronounced outer well minima correspond to $\omega_o = 2\pi~\,\cdot $~300~kHz.}\label{fig:tripleWellSchematic}
\end{figure}

Except for the harmonic case in Sec.~\ref{sec:CouplingInACommonWell} (see below), for the purpose of creating LDE, one can conceptually think of the ions being trapped in three wells: the center well confines the bulk ions, the coupling to the messenger ions in the outer wells can be varied by the well separation and the well curvatures. Three separate wells require a polynomial of at least 6th order to be modelled. If the potential has reflection symmetry with respect to the center of the middle well, only even powers remain and, neglecting a vertical offset, only three parameters specify the entire potential: the well separation $x_o$, and the curvatures at the center and at the outer wells, specified by the local trap frequencies $\omega_c$ and $\omega_o$, respectively, and the potential has the form
\begin{equation}
\phi(x) = m \left(\frac{2\,\omega_c^2+\omega_o^2}{12 x_o^4}x^6 -\frac{4\,\omega_c^2+\omega_o^2}{8 x_o^2}x^4+ \frac{\omega_c^2}{2}x^2 \right).
\label{eq:tripleWell}
\end{equation}
Figure~\ref{fig:tripleWellSchematic} shows the symmetric triple-well potential for constant well separation and central curvature and a variation of the trap frequency of the outer well $\omega_o$.

In the following, we discuss five potential shapes, and we analyze the resulting spin-spin coupling patterns for
the cases in which the trap is loaded with 4 or 6 ions.
Those are the simplest experimental situations in which the long distance entanglement can be observed and are the cases that we will analyze in detail in the remainder of this article.
The corresponding values for the ion positions, the qubit level splittings (for zero offset field), the normal modes and the spin-spin couplings are reported in the tables of appendix~\ref{sec:coupldata}.
Note that, for the mirror symmetry discussed above, the ion positions are symmetric (unless forced to be asymmetric by prior splitting and shuttling operations), and the couplings are also symmetric, specifically for a string of four ions
\[
J_{1,2} = J_{3,4} \quad\mbox{and}\quad  J_{1,3} = J_{2,4}.
\]
Our simulations slightly deviate from this symmetry (see, for example Tab.~\ref{table:LDE6} in appendix~\ref{sec:coupldata}), as we take into account the real geometry of our segmented trap. The relative deviations in the couplings from a symmetric pattern are on the order of a few percent, so for all four ion couplings this is neglected and only one value is given for each almost identical pair of couplings.

The lowest mode was kept at $\nu_0 = 2\pi \cdot 50$~kHz in all patterns, to produce scenarios with comparable susceptibility to finite temperature and stray fields. The maximum effective Lamb-Dicke parameter, as defined in \cite{Mintert2001,Mintert2003} was chosen to be $\eta_{\rm max}=0.1$ for all simulations to have comparable coupling between internal and motional states.

\subsubsection{Coupling in a common harmonic well}
\label{sec:CouplingInACommonWell}

\begin{figure}[ht]
\centering
\includegraphics[width=0.95\columnwidth]{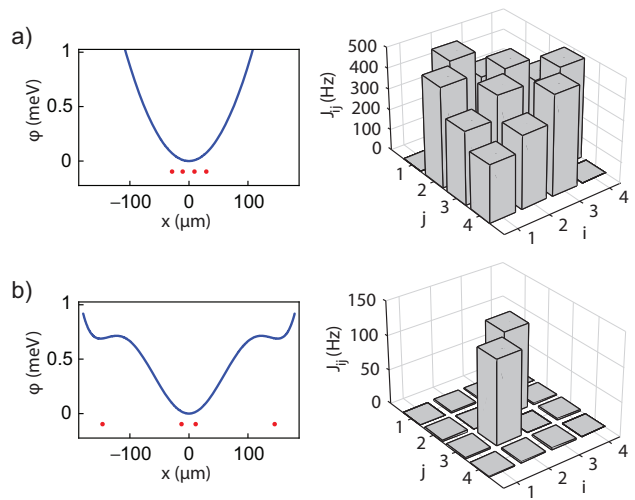}
\caption{a) Axial potential and equilibrium ion positions for four ions in a a predominantly harmonic single well. The values correspond to Tab.~\ref{table:harmonic} in appendix~\ref{sec:coupldata}.
\hspace{1mm}
b) Axial potential and equilibrium ion positions for four ions confined in three wells. The values correspond to Tab.~\ref{table:3well} in appendix~\ref{sec:coupldata}.}\label{fig:potCouplingFlatStraight}
\end{figure}

The first voltage pattern discussed here is optimized to give an almost purely quadratic dependence in the vicinity of its minimum. The normal modes are strongly delocalized and the coupling pattern shows next neighbor coupling but also long range couplings beyond next neighbors with almost identical strength. Fig.~\ref{fig:potCouplingFlatStraight}~a) shows the potential in the vicinity of its minimum together with the equilibrium positions and the resulting coupling pattern. Numerical values for positions, level splittings (compared to the situation with no gradient), normal modes and couplings are given in Tab.~\ref{table:harmonic} of appendix~\ref{sec:coupldata}.

\subsubsection{Coupling in three wells}
\label{sec:CouplingInThreeWells}
Trapping ions in three independent wells is an intuitive approach to generate LDE: the inner ions are confined in a common well and couple strongly. The outer ions are located in separate wells and show only small coupling to the center 'bulk' string due to their large separation and, depending on the shape of the outer well, potentially due to a stiff confinement (see Fig.~\ref{fig:potCouplingFlatStraight}~b), left image).
All eigenvectors corresponding to the normal modes are predominantly localized to one single well. Thus the outer ions couple weakly to all others which can be seen in the resulting coupling pattern (see Fig.~\ref{fig:potCouplingFlatStraight}~b), right image and Tab.~\ref{table:3well} in appendix~\ref{sec:coupldata}).  All couplings are negligible compared to the coupling between the two centre ions which are confined in the same well, and the situation is comparable to separate micro traps \cite{HWunderlich2009}. Numerical values for positions, levels splittings, normal modes and couplings are given in Tab.~\ref{table:3well} of appendix~\ref{sec:coupldata}.


\subsubsection{Coupling in a single strongly anharmonic well}
\label{sec:CouplingInASingleStronglyAnharmonicWell}

\begin{figure}[!t]
\centering
\includegraphics[width=0.95\columnwidth]{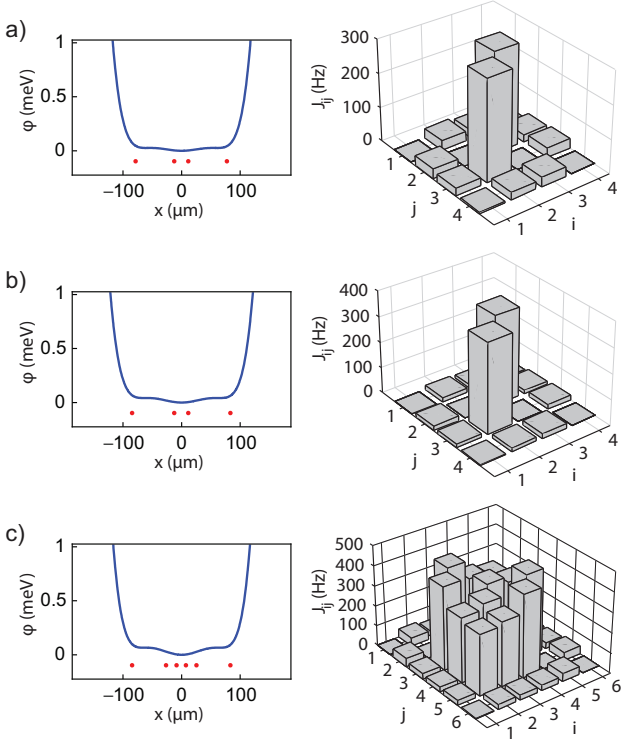}
\caption{Axial potential and equilibrium ion positions for four, a) and b), and six, c), ions confined in a single anharmonic well, whose shape generates a coupling pattern suitable for creating LDE.
Plots a), b) and c) correspond, respectively, to the values in Tabs.~\ref{table:LDE_0}, \ref{table:LDE} and \ref{table:LDE6} of appendix~\ref{sec:coupldata}.
}\label{fig:potCouplingLDE}
\end{figure}

Making a potential well strongly anharmonic, substantially alters ion positions \cite{Lin2009} and normal mode spectrum and allows to generate a pattern suitable for creating LDE. Changing $\omega_o$ allows to choose the ratio of the coupling of outer ions to their neighbors with respect to the coupling between the two center ions in a wide range
(see Fig.~\ref{fig:potCouplingLDE}). Note that three very shallow minima can be created within a region of 150~\textmu m, whereas the segment width of 130~\textmu m (of the trap that serves as a concrete example here) suggests that for a na\"{i}ve alternating voltage pattern three minima would have a spatial extent of approximately 500~\textmu m. Compared to ions trapped in three almost independent wells, where the two lowest modes are degenerate (the two outer ions oscillating separately), here there is one single lowest mode, separated from all others. The eigenvectors corresponding to the normal modes show stronger collective motion of all ions compared to ions trapped in individual wells.
The flatness of the potential over the region of the trapped ions indicates a sensitive dependence of the coupling on the applied voltages and puts strict requirements on voltage stability and accuracy which have to be taken into account in the design of the voltage supplies \cite{Baig2013}.
Figs.~\ref{fig:potCouplingLDE}~a) and b) correspond, respectively, to the two sets of numerical values for positions, levels splittings, normal modes and couplings given in Tabs.~\ref{table:LDE_0} and \ref{table:LDE} of appendix~\ref{sec:coupldata}.
The set in Tab.~\ref{table:LDE_0} corresponds to a slightly wider spatial configuration of the ions than that in Tab.~\ref{table:LDE}.

In the first case the resulting couplings of the outer ions is smaller hence the corresponding long distance entanglement is expected to be larger. On the other hand, the energy gap, that is, the energy difference between the eigenvalues of  $H_{\rm eff}$ corresponding to the ground and first excited states, for an XX spin model with such a set of coupling is smaller than that corresponding to the second set, and as a consequence the preparation time has to be larger in order for the adiabatic condition to be fulfilled during the dynamics (see App.~\ref{Adiab}).


A potential very similar to the one described above, optimized for six ions,
is shown in Fig.~\ref{fig:potCouplingLDE}~c). It is generated by only slightly modifying the potential shown in Fig.~\ref{fig:potCouplingLDE}~b), and keeping
 the softest mode at $\nu_0 = 50$~kHz and the maximum Lamb-Dicke paramete at $\eta_{\rm max} = 0.1$. The corresponding numerical values for positions, modes and couplings are reported in Tab.~\ref{table:LDE6} of appendix~\ref{sec:coupldata}.

\subsection{Spin dynamics with microwave pulses} \label{spindynamics}

The spin dynamics can be manipulated using external microwave fields that can drive selectively a given spin by tuning the driving frequency to the corresponding resonance \cite{Khromova2012}.
The hamiltonian for the interaction between the ion spins and the driving field in the rotating wave approximation and neglecting the ion motion (see Sec.~\ref{motion} for an analysis of the effects of the motion) takes the form
\begin{eqnarray}\label{HLt}
\overline H_L(t)&=&-\ii\hbar
\Omega(t) \sum_{j=1}^N\pg{\sigma_j^+ \ee^{-\ii \pq{\overline\omega(t) t +\varphi(t)}}-\sigma_j^- \ee^{\ii\pq{ \overline\omega(t) t +\varphi(t)}}}.
\end{eqnarray}
In general the amplitude $\Omega$, the frequency $\overline\omega$, and the phase $\varphi$ of the driving field can be time dependent.
In particular we consider a sequence of step-like driving pulses: We identify a set of time instants $t_m$ with $m=0,1,2,\cdots$ which define a corresponding set of time intervals $(t_{m-1}, t_m]$ (see lower part of Fig.~\ref{fig:pulsescheme}) during which  the driving Hamiltonian is constant,
($\Omega(t)=\Omega_m$, $\overline\omega(t)=\overline\omega_m$ and $\varphi(t)=\varphi_m$ for $t\in(t_{m-1}, t_m] $). In certain intervals the driving field can also be zero. If we define the square-pulse function
\begin{eqnarray}
\epsilon_m(t) =\theta(t-t_{m-1})-\theta(t-t_{m})
\end{eqnarray}
with  $\theta(t)=0$ for $t<0$ and $\theta(t)=1$ for $t\geq 0$, then Eq.~\rp{HLt} takes the form
\begin{eqnarray}
\overline H_L(t)&=&\sum_{m}\epsilon_m(t)\overline H\al{m}_L(t)
\end{eqnarray}
where
\begin{eqnarray}
\overline H_L\al{m}(t)&=&-\ii\hbar\Omega_m \sum_{j=1}^N\pq{\sigma_j^+ \ee^{-\ii\pt{ \overline\omega_m t +\varphi_m}}-h.c.}.
\end{eqnarray}
In each time interval in which $\Omega_m\neq0$, the driving frequency is close to resonance to a single spin $j_m$, with a small detuning $b_m=\omega_{j_m}-\overline\omega_m\ll\Omega_m$. All the other spins are far off resonance and their dynamics is not relevantly affected by the driving pulse.
Hence,
the Hamiltonian of the system, including the driving field reads
\begin{eqnarray}\label{barHt}
\overline H(t)&=&\overline H_z+ H_{zz}+\overline H_L(t)
\end{eqnarray}
with $\overline H_z$ and $H_{zz}$ defined in Eq.~\rp{barHIsing}.

The system dynamics is more conveniently  analyzed in a reference frame rotating at the driving field frequency as detailed in App.~\ref{refFrame}.
The Hamiltonian in  the new representation takes the form
\begin{eqnarray}\label{Ht}
H(t)&=&\sum_{m} \epsilon_m(t) \pq{ H_{z}\al{m}+H_{zz}+H_L\al{m}(t)}
\end{eqnarray}
with
\begin{eqnarray}\label{Hzm}
H_{z}\al{m}&=& \frac{\hbar}{2}b_m\sum_{j=1}^N\sigma_j^z
\\
H_L\al{m}(t)&=&-\ii\hbar\Omega_m \sum_{j=1}^N\pg{\sigma_j^+ \ee^{-\ii\Delta_{j_m,j}\,t}-h.c.},
\label{HLm}
\end{eqnarray}
%
%
where we have introduced the detuning between the spin resonance frequencies $$\Delta_{k,j}=\omega_k-\omega_j.$$
In this specific reference frame (see App.~\ref{refFrame} for details), the effective magnetic field along the $z-$axes, i.e. $b_m$ in Eq.~\rp{Hzm}, is the same for all spins. On the other hand, the spin-spin detuning which is much larger of both the effective magnetic field and the Rabi frequency, $\Delta_{j_m,j}\gg\Omega_m\gg b_m$, for $j\neq j_m$ , enters into the new time dependent driving Hamiltonian in Eq.~\rp{HLm}. In each time interval $m$, only the spin $j_m$ is driven resonantly ($\Delta_{j_m,j_m}=0$) , and it is the only spin that is relevantly affected by the driving field.
Correspondingly, the non resonant terms in Eq.~\rp{HLm} can be neglected and the Hamiltonian in Eq.~\rp{Ht} can be approximated as
\begin{eqnarray}\label{Ht0}
H(t)\simeq\hbar\sum_{m} \epsilon_m(t) \pq{\frac{b_m}{2}\sum_{j=1}^N  \sigma_j^z-\frac{1}{2}\sum_{i,j} J_{ij} \sigma_i^z\sigma_j^z+\Omega_m \sigma_{j_m}^y }
\end{eqnarray}
where in each time step a single spin $j_m$ sees an additional effective magnetic field directed along the $y-$axes.
We highlight that the dynamics in the  two representations are related by  a unitary and local transformation, thus the corresponding entanglement properties  are equal in the two representations.

%
%

\subsection{Stroboscopic engineering of the $XX$ spin dynamics}\label{Strobo}

\begin{figure}[ht]
\centering
\includegraphics[width=0.95\columnwidth]{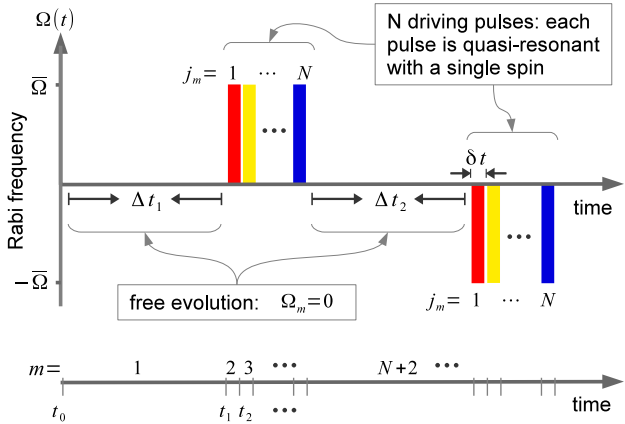}
\caption{Sequence of driving pulses corresponding to the evolution operator in Eq.~\rp{Ut}.}
\label{fig:pulsescheme}
\end{figure}

In order to engineer the dynamics of an $XX$ quantum spin model we consider a sequence of
driving pulses made of $2N+2$ steps ($m=1,\cdots 2N+2$), characterized by specific values of the parameters of the Hamiltonian in Eq.~\rp{Ht0}, as depicted in Fig.~\ref{fig:pulsescheme}. During the sequence of pulses the value of the detuning is fixed $b_m=b$, $\forall m$ .
While the $2N+2$ time steps are engineered as follows (see Fig.~\ref{fig:pulsescheme}):
after a free evolution (no driving) of time $\Delta t_1$, each spin is driven sequentially with a Rabi frequency $\overline\Omega$ and for a time $\delta t$ in order to realize $\pi/2$-pulses, i.e. 	 $\overline\Omega\ \delta t={\pi}/{4}$; then after another free evolution of time $\Delta t_2$, the spins are driven again sequentially with opposite phase (that is, in the second train of pulses the value of Rabi frequency is the opposite than that in the first train of pulses).
%
%
%
%

As discussed in App.~\ref{Sequence} the evolution operator corresponding to this sequence, at the final time $\bar t=\Delta t_1+\Delta t_2+N\, \delta t$,  with
\begin{eqnarray}\label{cond}
\overline\Omega\ \delta t&=&\frac{\pi}{4}
\end{eqnarray}
can be approximated, in the limit $\abs{\Delta_{j,j_m}}\gg \abs{\overline\Omega}\gg \abs{b_m},\abs{J_{j,k}}$,
as
\begin{eqnarray}\label{Ut}
U_{\bar t}= \ee^{-\ii H_{Ising}\al{x} \Delta t_2} \ \ee^{-\ii H_{Ising}\al{z} \Delta t_1}
\end{eqnarray}
where
\begin{eqnarray}
H_{Ising}\al{\zeta}&=& \frac{b}{2}\sum_{j=1}^N\sigma_j^\zeta- \frac{1}{2}\sum_{j,k} J_{ij} \sigma_i^\zeta\sigma_j^\zeta ,  \ \ \  {\rm for}\ \ \zeta\in\pg{x,z}.
\end{eqnarray}
Hence, the stroboscopic evolution at times $n \bar t$, with $n\in\mathbb{N}$, given by the repeated application of this sequence of pulses is described by the operator
\begin{eqnarray}\label{Unt}
U_{n\bar t}={U_{\bar t}}^n=\pt{\ee^{-\ii  H_{\rm Ising}\al{x} \Delta t_2}\ee^{-\ii H_{\rm Ising}\al{z}\Delta t_1}}^n.
\end{eqnarray}
According to the Trotter formula~\cite{Lloyd}
\begin{eqnarray}
\ee^{-i(H_1+H_2) t}=\lim_{n\to\infty}\pt{\ee^{-i H_1 t/n}\ee^{-i H_2 t/n}}^n,
\end{eqnarray}
and in the limit $\Delta t_1,\Delta t_2\ll \abs{J_{ij}}^{-1},\abs{b}^{-1}$, we can approximate the evolution operator in Eq.~\rp{Unt} as
\begin{eqnarray}
U_{n\bar t} \simeq \ee^{-\ii \pt{H_{\rm Ising}\al{z}+\alpha H_{\rm Ising}\al{x}} n\Delta t_1}=\ee^{-\ii \beta\pt{H_{\rm Ising}\al{z}+\alpha H_{\rm Ising}\al{x}} n\bar t}
\end{eqnarray}
where
 \begin{eqnarray}\label{alphabeta}
 \alpha&=&\frac{\Delta t_2}{\Delta t_1}
 \nn\\
\beta&=&\frac{\Delta t_1}{\bar t}=\frac{\Delta t_1}{(1+\alpha)\Delta t_1+2N\,\delta t}.
\end{eqnarray}
This result demonstrates that the stroboscopic evolution defined by Eq.~\rp{Unt} approximates the evolution, at times $n \bar t$, of a spin system with the effective Hamiltonian
\begin{eqnarray}\label{Heff}
H_{\rm eff}= \beta\pq{H_{\rm Ising}\al{z}+\alpha H_{\rm Ising}\al{x}}.
\end{eqnarray}

\section{Adiabatic preparation and stroboscopic/pulsed dynamics}\label{AdiabPulse}

\begin{figure*}[t!]
\centering
\includegraphics[width=18cm]{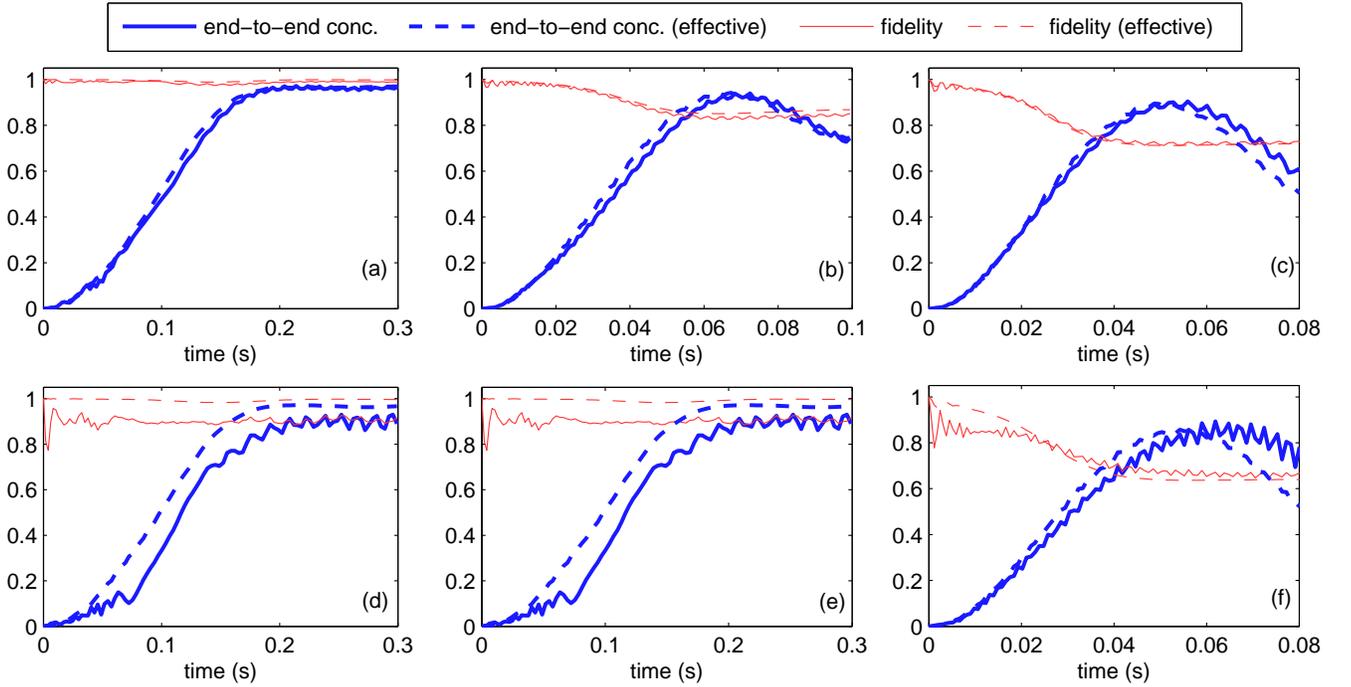}
\caption{End-to-end concurrence (thick, blue lines) and fidelity (thin, red lines) with the instantaneous ground state for a chain of four ions.
The spin-spin coupling constants and the spin resonance frequencies are identified in Tab.~\ref{table:LDE_0}. The dashed lines are obtained integrating the time dependent effective Hamiltonian in Eq.~\rp{Heff} with the time dependence defined in Eq.~\rp{alphah} ($b_0=2\pi\times 0.1$~kHz); The solid lines are obtained using the Hamiltonian in Eq.~\rp{Ht0}, following the pulse scheme described in Sec.~\ref{Strobo} and with the stepwise variation of the parameters $b$ and $\alpha$.
The upper plots, (a), (b) and (c) are obtained with  ${\delta t}=1$~\textmu s as defined in Eq.~\ref{cond} and the lower plots (d), (e) and (f) with ${\delta t}=5$~\textmu s.
From left to right the velocity of the adiabatic manipulation is gradually increased: in (a), (d) $r=2\pi\times 3.2$~Hz, in (b), (e) $r=2\pi\times 10.6$~Hz, and in (c), (f) $r=2\pi\times 15.9$~Hz.
In all plots  $\Delta t_1=100\,$~\textmu s.}
\label{fig:results1_a}
\end{figure*}

\begin{figure*}[t!]
\centering
\includegraphics[width=18cm]{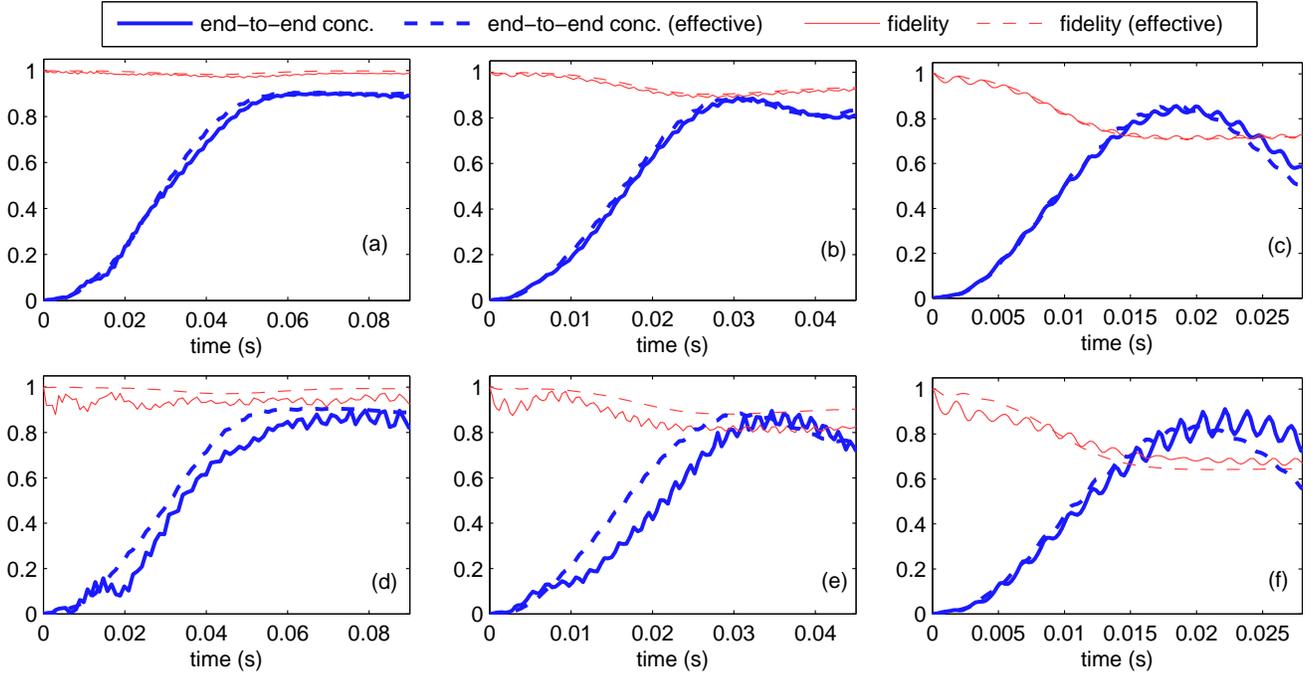}
\caption{As in Fig.~\ref{fig:results1_a} with the spin-spin coupling constants and the spin resonance frequencies defined in Tab.~\ref{table:LDE} and with ${\delta t}=1$~\textmu s. Moreover in (a) $r=2\pi\times 10$~Hz, in (b) $r=2\pi\times 20$~Hz, and in (c) $r=2\pi\times 40$~Hz.
}
\label{fig:results1}
\end{figure*}

\begin{figure}[t!]
\centering
\includegraphics[width=8.5cm]{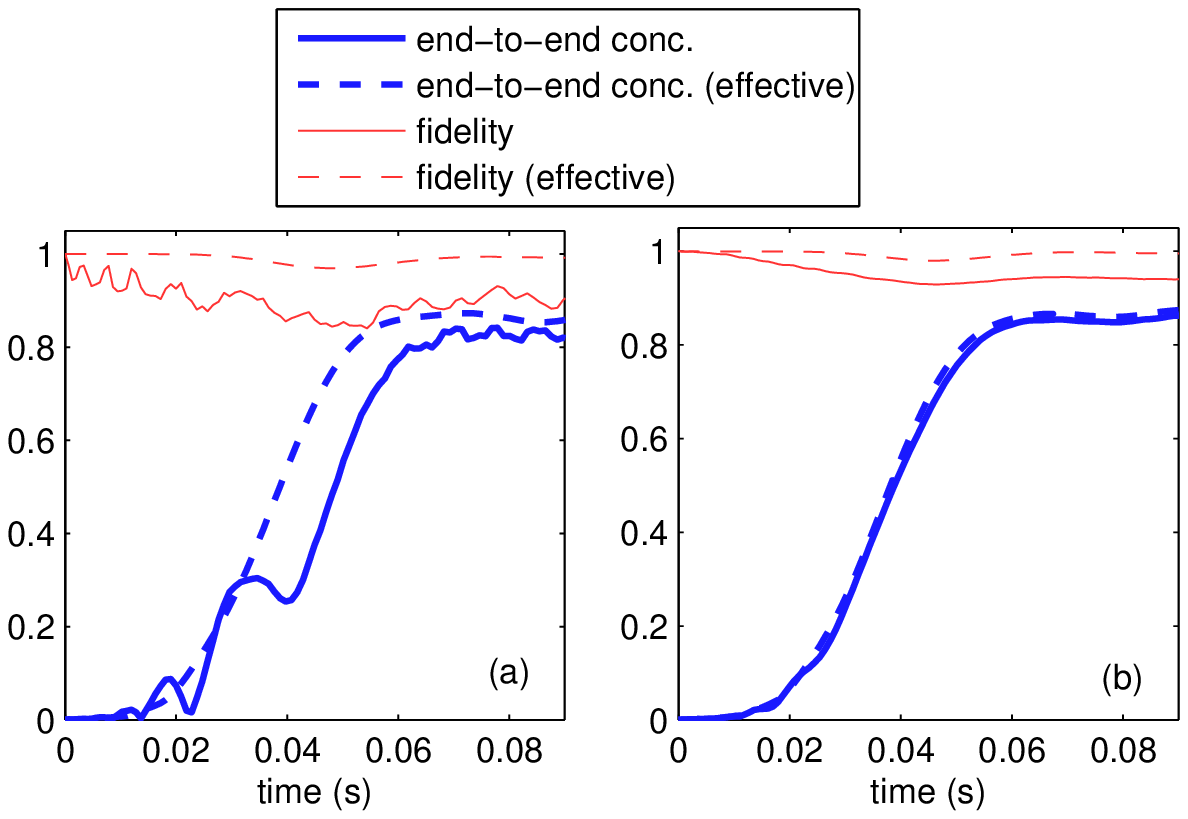}
\caption{Results obtained for a chain of six ions. The lines code is as in Fig.~\ref{fig:results1}.
The spin-spin coupling matrix is reported in Tab.~\ref{table:LDE6}.
In (a) ${\delta t}=1$~\textmu s, while in (b) ${\delta t}=0.5$~\textmu s.
The other parameters are $b_0=2\pi\times 0.2$~kHz, $r=2\pi\times 8$~Hz, and $\Delta t_1=40$~\textmu s.
}
\label{fig:results1c}
\end{figure}

The parameters $b$ and $\alpha$ (and consequently $\beta$) can be varied adiabatically in order to prepare the ground state of an $XX$ hamiltonian: The effective external magnetic field $b$  is varied by adjusting the detuning between the driving field and the ion-spin resonance frequencies. On the other hand $\alpha$, and correspondingly $\beta$ are varied by controlling the time $\Delta t_2$ (see App.~\ref{Adiab} for details). If the variation is slow enough, then the system will follow adiabatically the instantaneous ground state of the effective hamiltonian.

The system is initialized in the ferromagnetic state with all the spins aligned along the $z$-axes, that is, the ground state of the Ising Hamiltonian ($\alpha=0$). The value of $b$ is initially set to some value $b_0$ in order to remove the ground state degeneracy.

The parameters $\alpha$ and $b$ are then slowly varied to realize the adiabatic preparation of the LDE. In particular during each sequence of pulses, that is described in Sec.~\ref{Strobo}, the values of $b$ and $\alpha$ are kept fixed, while they are varied from sequence to sequence in order to realize a stepwise approximation of the functions (see App.~\ref{Adiab})
\begin{eqnarray}\label{alphah}
\alpha(t)&=&1-\ee^{-r t}
\nn\\
b(t)&=&b_0\ee^{-r t},
\end{eqnarray}
where $r$ is the rate of variation of the hamiltonian. By these means at large time $b=0$ and $\alpha=1$ so that the final effective Hamiltonian  is of $XX$ type. In particular  for sufficiently small $r$ the corresponding ground state is achieved.

The efficiency of this
stepwise adiabatic protocol is analyzed numerically by evaluating the evolution corresponding to the Hamiltonian~\rp{Ht} with the time sequence and the parameters discussed in Sec.~\ref{Strobo}, and the corresponding stepwise variation of $b$ and $\alpha$.
The results are shown in Figs.~\ref{fig:results1_a}
(solid lines) for different values of $r$ and using the parameters reported in Tab.~\ref{table:LDE_0}. They are compared with that obtained by the numerical integration of the Schr\"odinger equation with the effective time dependent Hamiltonian in Eq.~\rp{Heff} where the time dependent parameters $\alpha$ and $b$ are defined in Eq.~\rp{alphah} (dashed lines).

The protocol is characterized in terms of the fidelity between the resulting state and the expected instantaneous ground state of the effective Hamiltonian~\rp{Heff} (red, thin curves), and in terms of the end-to-end  concurrence (blue, thick curves). The fidelity indicates the extent to which the resulting state differs from the expected one: Fidelity equal to one corresponds to perfect adiabatic following; while
equal fidelity for both the standard adiabatic evolution (dashed lines) and the stepwise adiabatic evolution (solid lines) means that the protocol realizes a perfect simulation of the effective Hamiltonian. On the other end, the concurrence measures the entanglement between the end spins, and concurrence equal to one indicates a maximally entangled Bell state.

When the rate of variation of the hamiltonian parameters $r$ is sufficiently small (Fig.~\ref{fig:results1_a}) (a)  the ground state preparation is good: The fidelity is close to one and as expected the ground state exhibits large entanglement. When on the other hand the rate is increased (Figs.~\ref{fig:results1_a} (b)  and (c)) then the evolution  is no more adiabatic and the system ends up in a state which is  not exactly the ground state of the final Hamiltonian and the fidelity is reduced. The faster the manipulation, the smaller is the corresponding fidelity. Nevertheless in all cases, the end-to-end entanglement can be very large at certain times  meaning that the end spins approach a Bell state.

In all cases the results obtained with the effective Hamiltonian and that obtained via the sequence of pulses are similar meaning that the protocol is faithful and a good simulation of the effective model is realized.

Figs.~\ref{fig:results1_a} (d), (e) and (f) describe how the efficiency of the scheme is reduced when implemented with not sufficiently fast $\pi/2$ driving pulses. In this case during the pulses the system dynamics is not negligible  and the transformation which generate the spin-spin interaction along the $x$-axes is not exact.

%

Faster preparation of the LDE can be achieved with systems with larger gap. This can be obtained by careful shaping of the trapping potential as discussed in Sec.~\ref{sec:TailoringCouplingConstantsInASegmentedTrap}. Simulations realized with the spin-spin coupling strengths reported in Tab.~\ref{table:LDE} are shown in Fig.~\ref{fig:results1}. Here the preparation time is shorter than that of Fig.~\ref{fig:results1_a}.

Similar results are obtained also with larger chains, see Fig.~\ref{fig:results1c} that is realized with six ions. In this case  the driving pulses have to be made shorter in order to optimize the preparation as described by Fig.~ \ref{fig:results1c} (b).

\subsection{Effect of non-resonant spins}

The results that we have discussed so far are based on Eq.~\rp{Ht0} where we have neglected the effect of the driving field on the spins which are not close to resonance. This is justified when the difference in frequency between the spin resonances is much larger than the Rabi frequency $\Omega$: $\Delta_{j_m,j}\gg\Omega_m$. In this case the spins experience a dynamical Zeeman shift $\delta_{j}\al{m}$ whose magnitude can be evaluated in perturbation theory and is given by
\begin{eqnarray}
\delta\omega_{j_m,j}\simeq \abs{\frac{\Omega_m^2}{2\Delta_{j_m,j}}}.
\end{eqnarray}
Correspondingly, during a driving pulse on spin $j_m$, whose duration is $\delta t=\pi/4\overline\Omega$ the phase accumulated by spin $j$ as a result of the dynamical Zeeman shift is $\Phi_{j_m,j}=\delta t\times\delta\omega_{j_m,j}=\pi\abs{\overline\Omega/8\Delta_{j_m,j}}$.

The largest phase for the parameters of Figs.~\ref{fig:results1_a} and \ref{fig:results1} is $\Phi_{2,3}\simeq 7.6\times 10^{-3}$. Similarly we find that the largest phase for the parameters of Fig.~\rp{fig:results1c} is $\Phi_{3,4}=19.7\times10^{-3}$.
In all cases these values are very small and they justify our approximation.

\subsection{Mechanical effects}\label{motion}

So far we have neglected the motion of the ions. Internal electronic dynamics and motion can be coupled by an electromagnetic field. In particular when the ions are in a magnetic gradient also long wavelength radiation, as microwaves, can have a significant mechanical effect allowing for example for sideband cooling \cite{Mintert2001,Wunderlich2002,Wunderlich2005,Scharfenberger2013}.
In the following we justify our treatment in which we neglect the atomic motion.

In a magnetic gradient
the coupling between an ion $j$ and a mechanical normal mode $k$ is scaled by the effective Lamb-Dicke (LD) parameters~\cite{Wunderlich2002}
\begin{eqnarray}
\eta_{j,k}=\sqrt{\frac{\hbar}{2 m\, \nu_k}}\frac{\mu_B\ g}{\hbar\, \nu_k} \rbar{\frac{\partial B}{\partial x}}_{x=x_{0,j}} S_{j,k}
\end{eqnarray}
where $\nu_k$ is the frequency of the normal modes, and  $S$ is the matrix that diagonalize the Hessian matrix $A$ (see Sec.~\ref{system}) of the potential energy function that confine the ions, that is $\pt{S^TAS}_{j,k}=\delta_{j,k}\ m\, \nu_j^2$.
These parameters are typically small and allow for a systematic expansion of the corresponding dynamics in power of $\eta_{j,k}$.
Including the lowest order mechanical effects the Hamiltonian for the interaction between the ions and the driving field (see also Eq.~\rp{HLt}) takes the form
\begin{eqnarray}
\overline H_L(t)&=&-\ii\hbar
\Omega(t)
\\&&\times \sum_j
\pg{\sigma_j^+\pq{1+\sum_k\eta_{j,k}\pt{a\da_k-a_k}} \ee^{-\ii \pq{\overline\omega(t) t +\varphi(t)}}
-h.c.}\nn
\end{eqnarray}
where $h.c.$ stands for the hermitian conjugate, and $a_k\da$, $a_k$ are the creation and annihilation operators for the vibrational mode $k$.
This Hamiltonian accounts for sideband transitions at frequencies $\omega_j\pm\nu_k$.
The mechanical transitions are negligible when
\begin{eqnarray}
\eta^2\ \pt{\overline n_k+1}\ll1
\end{eqnarray}
 where $\overline n_k$ is the average number of excitations in the the vibrational mode $k$.

For the parameters used in Figs.~\ref{fig:results1_a}, \ref{fig:results1} and \ref{fig:results1c},
the LD parameters take values between $0.1$ and $2.5\times 10^{-6}$,
which demonstrate the validity of our results also for Doppler cooled trapped ions, without additional sub Doppler cooling to the ground state of the axial potential.
Increasing the gradient of the $B$ field, the coupling strengths increases allowing for a faster preparation; However the system approaches the regime in which the mechanical effects are relevant.
In fact, stronger gradient of the magnetic field corresponds to larger LD parameters.

\subsection{Effect of spin dephasing}

\begin{figure*}[t!]
\centering
\includegraphics[width=18cm]{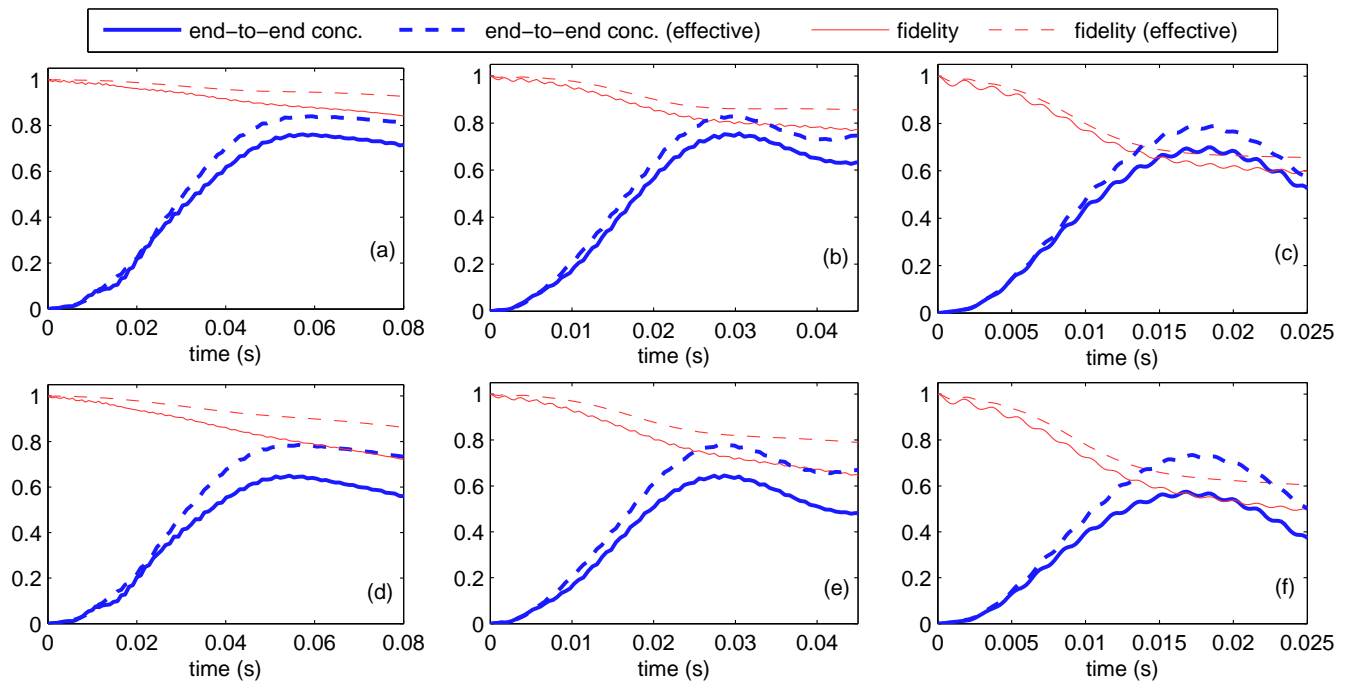}
\caption{
Plots (a), (b) and (c):
as Figs.~\ref{fig:results1} with dephasing time $T_{deph}\equiv 1/\gamma=50/ r$, that is in (a) $T_{deph}=0.8$~s, in (b) $T_{deph}=0.4$~s, and in (c) $T_{deph}=0.2$~s.
Plots (d), (e) and (f):
as Figs.~\ref{fig:results1} with dephasing time $T_{deph}=25/ r$, that is in (d) $T_{deph}=0.4$~s, in (e) $T_{deph}=0.2$~s, and in (f) $T_{deph}=0.1$~s.
}
\label{fig:results2}
\end{figure*}

%
%

In practice stray magnetic fields induce fluctuating spin resonance frequencies, which in turn induce decay of the spin coherence, namely depahsing.
The curves in Fig.~\ref{fig:results2}
are evaluated including the dephasing of the spins.
They are obtained by solving a master equation for the spins  dynamics of the form
\begin{eqnarray}
\dot\rho=-\ii\pq{H(t),\rho}+\LL_D\rho
\end{eqnarray}
where $H(t)$ corresponds to the Hamiltonian~\rp{Ht0} for the solid lines and to the effective Hamiltonian~\rp{Heff} for the dashed lines. Moreover $\LL_D$ accounts for the spins dephasing at rate $\gamma$
and takes the form
\begin{eqnarray}
\LL_D\rho=\frac{\gamma}{2} \sum_j\pt{\sigma_j^z\ \rho\ \sigma_j^z-\rho}.
\end{eqnarray}
This model describes a system of spins with randomly fluctuating resonance frequencies $\omega_j=\omega_{j,0}+\xi_j(t)$, where $\xi_j(t)$ are delta-correlated random variables (i.e. $\av{\xi_j(t),\xi_j(t')}\propto \delta(t-t')$).
As expected, the dephasing reduces the efficiency of the scheme, and both the resulting fidelity and concurrence are slightly lower then the corresponding ones obtained without dephasing.
In particular, Fig.~\ref{fig:results2} shows that the scheme works also under the effect of dephasing processes with dephasing times sufficiently larger than the preparation time. Where the preparation time is roughly one order of magnitude larger than $1/r$ with $r$ the rate of variation of the Hamiltonian parameters which is  introduced in Eq.~\rp{alphah}.

Resistance to dephasing can be achieved by implementing
dynamical decoupling techniques~\cite{Piltz2013}. In fact the Trotter expansion scheme makes use of microwave pulses for rotating the Bloch vector by $\pi/2$ to achieve a 	stroboscopic implementation of $XX$-Hamiltonians. Hence, by changing phases of all microwave pulses by $\pi$ at every other instance, the system effectively flips for example between $z$, $x$, $-z$, and $-x$ and refocussing of low frequency noise components is achieved at no extra cost.

\subsection{Experimental Feasibility}
\label{sec:ExperimentalFeasibility}

In Ref. \cite{Kaufmann2011} the operation of a  segmented trap with a built-in switchable gradient based on micro-structured solenoid is reported. Different isotopes of Ytterbium with or without hyperfine structure can be trapped. For the experiment proposed here, we use \ybu\  with a nuclear spin of $I=1/2$ yielding two hyperfine levels with $F=0, F=1$ in the electronic ground state \cite{Timoney2008,Khromova2012}. Different qubit implementations are possible: either magnetic sensitive states can be used to allow for \textbf{ma}gnetic \textbf{g}radient \textbf{i}nduced \textbf{c}oupling (MAGIC) \cite{Mintert2001,Wunderlich2002,Johanning2009} as required for the experiments discussed here. Or magnetic insensitive states can be used to yield a quantum memory with a long coherence time\cite{Wunderlich2003}.

In this segmented trap experiment \cite{Kaufmann2011} ,  the qubit is manipulated using microwave fields (applied through a conventional wave guide) and Rabi frequencies exceeding $\Omega \approx 100$~kHz on the $\sigma$ transition and a bare coherence time of the magnetic field sensitive hyperfine qubit on the order of 5~ms have been observed. Applying spin echo techniques \cite{Hahn1950}, dynamical decoupling \cite{Piltz2013} or dressed states \cite{Timoney2011}, we expect to be able to observe a coherent time evolution on a second timescale. The gradients required for the experiments here are expected to be possible with the present setup.
Splitting and merging of ion strings (which involves the generation of anharmonic trapping potentials), as well as shuttling have been demonstrated. Stable trapping could be observed down to axial trap frequencies around $\nu_1 = 2\pi \cdot 40$~kHz.

In order to improve the level of control over the anharmonicity of the axial trapping potential, it might be necessary to use smaller axial trapping segments, possibly in a surface trap. Larger gradients would boost the coupling and allow for stiffer axial confinement, making the scheme more robust against thermal excitation and external stray fields.

\begin{figure}[ht]
\centering
\includegraphics[width=0.95\columnwidth]{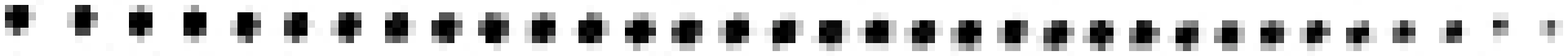}
\caption{Chain of 33 \ybg~ions in our segmented trap.}\label{fig:chain33}
\end{figure}

\section{Conclusion and Outlook}
\label{sec:Outlook}

In this article we have introduced and investigated schemes for the the implementation of LDE with trapped ions. The spin-spin Hamiltonians required for this purpose may as well be used for quantum simulations. In particular, we have shown how to tailor the trapping potential in order to engineer a specific spin-spin coupling pattern in one-dimensional lattices, and we have designed a sequence of microwave pulses able to engineer effective spin-1/2 Hamiltonians of $XX$ type. The same technique can be used to engineer any kind of isotropic and anisotropic Heisenberg and $XY$ models. In this perspective, our scheme and techniques may be expected to be especially useful for the verification of recent predictions about some nontrivial ground-state entanglement patterns, including field-interaction balancing and the onset of ground-state factorization  \cite{GSF2008,GSF2009}; general bounds between universal measures of frustration and ground-state entanglement \cite{Frustration2011,Frustration2013}; and universality in the scaling behavior of the entanglement spectrum \cite{ScalingEntSpectrum2013}. Finally, we have analyzed the efficiency of the adiabatic quantum preparation of the ground state of an effective Hamiltonian which exhibits LDE, demonstrating its feasibility within the limits of current ion trap technology. In the course of the investigation, we have introduced and combined for the first time trap shaping, adiabatic preparation, and Trotterization of the interactions, three elements that have not been perviously combined together. These elements are necessary for the realization, so far not yet attained, of highly nonclassical features of complex models of interacting quantum many-body systems.

\section*{Acknowledgments}
We thank Andr\'{e}s F. Var\'{o}n for fruitful discussions. We acknowledge funding by the European Community's Seventh Framework Programme (FP7/2007-2013) under grant agreements number 270843 (iQIT) and number 249958 (PICC) and by Bundesministerium f\"{u}r Bildung und Forschung (FK 01BQ1012). SMG acknowledges the Austrian Science Found (FWF-P23627-N16).

\newpage
\appendix
\section{Ion positions, normal modes, transition frequencies and couplings}\label{sec:coupldata}

\begin{table}[ht!]
\begin{center}
	\begin{tabular}[t]{ccccc}
i & 1 & 2 & 3 & 4\\
$x_i$ (\textmu m) & -28.7 & -8.9 & 8.3 & 29.3 \\
$\Delta\omega_i / 2\pi$ (MHz) & -11.8 & -3.7 & 3.9 & 12.0\\
$\nu_i / 2\pi$ (kHz) & 50.0 & 86.6 & 120.5 & 152.6\\
\\
$i,j$ &  1,2 & 1,3 & 1,4 & 2,3\\
$J_{ij}$ (Hz) & 479 & 349 & 273 & 457
	\end{tabular}
\end{center}
\caption{Positions $x_i$, changed qubit splittings $\Delta\omega_i$, normal modes $\nu_i$ and couplings $J_{ij}$ for an ion chain of four ions in an approximately harmonic axial trapping potential. The gradient required to obtain this values is 29.38~T/m.}
\label{table:harmonic}
\end{table}

\begin{table}[ht!]
\begin{center}
	\begin{tabular}[t]{ccccc}
i & 1 & 2 & 3 & 4\\
$x_{0,i}$ (\textmu m) & -145.9 & -10.8 & 11.4 & 146.4\\
$\Delta\omega_i / 2\pi$ (MHz) & -30.7 & -2.3 & 2.4 & 30.9\\
$\nu_i / 2\pi$ (kHz) & 50.0 & 50.1 & 59.9 & 105.4\\
\\
$i,j$ &  1,2 & 1,3 & 1,4 & 2,3\\
$J_{ij}/2\pi$ (Hz) & 2.1 & 1.8 & 0.4 & 123.8
\end{tabular}
\end{center}
\caption{Positions $x_i$, qubit splittings $\omega_i$, mechanical normal modes $\nu_i$ and couplings $J_{ij}$ for an ion chain of four ions in three approximately harmonic axial trapping potential. The gradient required to obtain this values is 15.06~T/m.}
\label{table:3well}
\end{table}

\begin{table}[ht!]
\begin{center}
	\begin{tabular}[t]{ccccc}
i & 1 & 2 & 3 & 4\\
$x_{0,i}$ (\textmu m) & -83.0 & -11.8 & 12.3 & 83.7\\
$\Delta\omega_i / 2\pi$ (MHz) & -24.2 & -3.4 & 3.6 & 24.4\\
$\nu_i / 2\pi$ (kHz) & 50.0 & 71.8 & 72.3 & 91.8\\
\\
$i,j$ &  1,2 & 1,3 & 1,4 & 2,3\\
$J_{ij}/2\pi$ (Hz) & 17.6 & 13.8 & 1.3 & 351.5
\end{tabular}
\end{center}
\caption{Positions $x_i$, qubit splittings $\omega_i$, normal modes $\nu_i$ and couplings $J_{ij}$ for an ion chain of four ions in a strongly anharmonic axial trapping potential. The gradient required to obtain this values is 20.81~T/m.}
\label{table:LDE_0}
\end{table}
\begin{table}[ht!]
\begin{center}
	\begin{tabular}[t]{ccccc}
i & 1 & 2 & 3 & 4\\
$x_{0,i}$ (\textmu m) & -76.9 & -11.7 & 12.2 & 77.6\\
$\Delta\omega_i / 2\pi$ (MHz) & -20.7 & -3.2 & 3.3 & 20.9\\
$\nu_i / 2\pi$ (kHz) & 50.0 & 59.1 & 59.7 & 92.4\\
\\
$i,j$ &  1,2 & 1,3 & 1,4 & 2,3\\
$J_{ij}/2\pi$ (Hz) & 28.8 & 22.2 & 2.2 & 298.8
\end{tabular}
\end{center}
\caption{Positions $x_i$, qubit splittings $\omega_i$, normal modes $\nu_i$ and couplings $J_{ij}$ for an ion chain of four ions in a strongly anharmonic axial trapping potential. The gradient required to obtain this values is 19.27~T/m.}
\label{table:LDE}
\end{table}

\begin{table}[ht!]
\begin{center}
	\begin{tabular}[t]{ccccccc}
i & 1 & 2 & 3 & 4 & 5 & 6\\
$x_{0,i}$ (\textmu m) & -84.2 & -26.0 & -7.8 & 8.3 & 26.4 & 84.9\\
$\Delta\omega_i / 2\pi$ (MHz) & -32.1 & -9.9 & -3.0 & 3.2 & 10.0 & 32.3\\
$\nu_i / 2\pi$ (kHz) & 50.0 & 90.0 & 90.9 & 92.1 & 136.2 & 181.7\\
\end{tabular}
\[
J_{ij} {\rm (Hz)}=
\left(
\begin{array}{cccccc}
0.&	27.9&	19.5&	16.7&	16.7&	1.4\\
27.9&	0.&	411.8&	319.7&	300.3&	16.5\\
19.5&	411.8&	0.&	348.3&	319.2&	16.4\\
16.7&	319.7&	348.3&	0.&	410.9&	19.1\\
16.7&	300.3&	319.2&	410.9&	0.&	27.3\\
1.4&	16.4794&	16.4&	19.1&	27.3&	0.
\end{array}
\right)
\]
\end{center}
\caption{Positions $x_i$, qubit splittings $\omega_i$, normal modes $\nu_i$ and couplings $J_{ij}$ for an ion chain of six ions in a strongly anharmonic axial trapping potential. The gradient required to obtain this values is 27.22~T/m.}
\label{table:LDE6}
\end{table}

\section{The dynamics in a new reference frame}\label{refFrame}

Let us consider the model described by the Hamiltonian in  Eq.~\rp{barHt},
and study the dynamics in a new reference frame defined by the unitary transformation
\begin{eqnarray}
U_0(t)&=&\sum_m\epsilon_m(t)\ee^{-\frac{\ii}{\hbar} H_0\al{m}\pt{t-t_{m-1}}}\ee^{-\frac{\ii}{\hbar} H_0\al{m-1}\pt{t_{m-1}-t_{m-2}}}\cdots \ee^{-\frac{\ii}{\hbar} H_0\al{1}\pt{t_{1}-t_{0}}}
\nn\\
&&+\theta(t_0-t)
\end{eqnarray}
with
\begin{eqnarray}
H_0\al{m}=\frac{\hbar}{2}\sum_j (\omega_j-b_m)\,\sigma_j^z
\end{eqnarray}
where $b_m$ is the detuning between the driving field frequency, $\overline\omega_m$, and the resonance frequency, $\omega_{j_m}$, of spin $j_m$ which is driven close to resonance in each time step, $b_m=\omega_{j_m}-\overline\omega_m$.
Since the unitary transformation is local, the entanglement properties in the new representation are the same as that in the original one.

If $\ke{\overline\psi(t)}$ is the state in the original representation, then the dynamics   of the transformed state $\ke{\psi(t)}=U_0\da(t)\ke{\overline\psi(t)}$ is ruled by the
Hamiltonian
\begin{eqnarray}
H(t)&=&U_0\da(t)\overline H(t)U_0(t)-\sum_m \epsilon_m(t)H_0\al{m}
\nn\\
&=&\sum_m\epsilon_m(t)\pq{H_{z}\al{m}+ H_{zz}+ H_L\al{m}(t)}
\end{eqnarray}
with
\begin{eqnarray}
H_{z}\al{m}=\overline H_{z}\al{m}-H_0\al{m} =\frac{\hbar}{2}\sum_j b_m\sigma_j^z
\nn\\
 H_L\al{m}=-\hbar\,\ii\, \Omega_m \sum_j\pg{\sigma_j^+ \ee^{-\ii\pq{\pt{\omega_{j_m}-\omega_j} t+\phi_m+\varphi_m}}-h.c.}
\end{eqnarray}
where
\begin{eqnarray}
\phi_m=\sum_{m'=1}^{m-1}\pt{b_{m'+1}-b_{m'}}t_{m'}.
\end{eqnarray}
The last Hamiltonian is obtained exploiting the relation $\ee^{\ii \zeta\sigma_j^z t}\sigma_j^+\ee^{-\ii \zeta\sigma_j^z t}=\sigma_j^+\ee^{2\ii \zeta t}$.
If the value of the phase of the driving field is set to the value
\begin{eqnarray}
\varphi_m=-\phi_{j_m}\al{m}
\end{eqnarray}
then
\begin{eqnarray}
 H_L\al{m}=-\hbar\,\ii\,\Omega_m \sum_j\pq{\sigma_j^+ \ee^{-\ii\pt{\omega_{j_m}-\omega_j} t}-h.c.}.
\end{eqnarray}
Thereby we obtain Eq.~\rp{Ht}.
We note that, in this representation, in each time step, the spin $j_m$ sees an effective magnetic field along the $y$-axes (see Eq.\rp{HLjm}).

\section{The sequence of driving pulses}\label{Sequence}

We are interested in the limit in which
$\abs{\omega_j-\omega_{j_m}}\gg \abs{\Omega_m}\gg \abs{b_m},\abs{J_{j,k}}$, for $j\neq j_m$.
Hence we can approximate the Hamiltonian~\rp{Ht} by retaining only the resonant terms as
\begin{eqnarray}\label{ApproxH}
H(t)&\simeq&\sum_{m\ :\ \Omega_m=0} \epsilon_m(t) \pq{ H_{z}\al{m}+H_{zz}}
\nn\\&&
+\sum_{m\ :\ \Omega_m\neq 0} \epsilon_m(t) H_L\al{m}
\end{eqnarray}
with
\begin{eqnarray}\label{HLjm}
H_L\al{m}&\simeq&\hbar\Omega_m\sigma_{j_m}^y.
\end{eqnarray}
where the sum over the time intervals is divided into two sums over the intervals in which the driving field is on ($\Omega_m\neq 0$) and off ($\Omega_m=0$) respectively.

The evolution operator corresponding to the sequence of pulses described in Sec.~\ref{Strobo} can be written, using the approximate Hamiltonian~\rp{ApproxH}, in the form
\begin{eqnarray}
U_{\bar t}&=&\ee^{-\ii H_N\al{+} \delta t}\cdots \ee^{-\ii H_1\al{+} \delta t}\  \ee^{-\ii H_{\rm Ising}\al{z} \Delta t_2} \ \ee^{-\ii H_1\al{-} \delta t}\cdots \ee^{-\ii H_N\al{-} \delta t}\
\nn\\&&\times\
\ee^{-\ii H_{\rm Ising}\al{z} \Delta t_1}
\nn\\
\end{eqnarray}
where
\begin{eqnarray}
H_{\rm Ising}\al{z}&=& \frac{b}{2}\sum_{j=1}^N\sigma_j^z- \frac{1}{2}\sum_{i,j} J_{ij} \sigma_i^z\sigma_j^z
\nn\\
H_{j_m}\al{\pm}&=&\pm\overline\Omega\,\sigma_{j_m}^y \ ,
\end{eqnarray}
and the total time of the sequence is
\begin{eqnarray}
\bar t=\Delta t_1+\Delta t_2+N\, \delta t.
\end{eqnarray}

Now we use the relation
\begin{eqnarray}
\Xi\pt{\Gamma}\equiv\ee^{-\ii \Gamma \sigma_j^y}\sigma_j^z\ee^{\ii \Gamma \sigma_j^y}
=\cos\pt{2\Gamma}\sigma_j^z+\sin\pt{2\Gamma}\sigma_j^x\ ,\nn
\end{eqnarray}
which reduces to $\Xi\pt{\Gamma}=\sigma_j^x$ when
$\Gamma=\frac{\pi}{4}+n\pi$ with $n\in \mathbb{Z}$.
Thus setting, for example,
\begin{eqnarray}\label{cond1}
\overline\Omega\, \delta t&=&\frac{\pi}{4}\ ,
\end{eqnarray}
 then
\begin{eqnarray}
U_{\bar t}= \ee^{-\ii H_{\rm Ising}\al{x} \Delta t_2} \ \ee^{-\ii H_{\rm Ising}\al{z} \Delta t_1}
\nn\\
\end{eqnarray}
where
\begin{eqnarray}
H_{\rm Ising}\al{x}&=& \frac{b}{2}\sum_{j=1}^N\sigma_j^x- \frac{1}{2}\sum_{i,j} J_{ij} \sigma_i^x\sigma_j^x.
\end{eqnarray}

\section{Adiabatic preparation of the ground state}\label{Adiab}

\begin{figure}[t!]
\centering
\includegraphics[width=0.9\columnwidth]{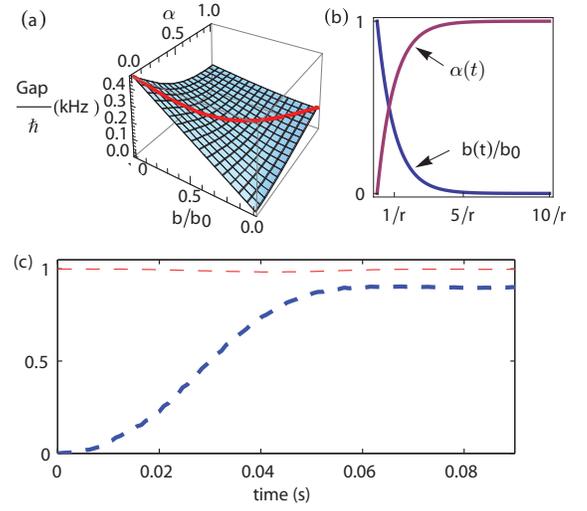}
\caption{(a) Gap between ground a first excited state of $H_{\rm eff}$ in Eq.~\rp{Heff} with four ions, in the space of parameters $\pg{b,\alpha}$. The red line indicates the gap corresponding to the adiabatic variation of $\alpha$ and $b$. (b) time evolution of the parameters $\alpha(t)$ and $b(t)$. (c) End-to-end concurrence (thick, blue line) and fidelity with the instantaneous ground state (thin, red line), obtained integrating the time dependent Schr\"odinger equation with the effective Hamiltonian in Eq.~\rp{Heff} with four ions, and with $b_0=2\pi\times 0.1$~kHz and $r=2\pi\times 10$~ Hz. The spin-spin couplings are reported in Tab.~\ref{table:LDE}.
Although not relevant for the present result, in order to be consistent with the results of Sec.~\ref{AdiabPulse}, we have set the parameter $\beta$ to the values defined in Eq.~\rp{alphabeta} (a different value of $\beta$ corresponds to a rescaling of the energy an	d correspondingly of the duration of the protocol). The two curves are equal to the dashed curves of Fig.~\ref{fig:results1} (a).}
\label{fig:adiab}
\end{figure}

A system initially in an eigenstate $\ke{\psi_j(0)}$ of its Hamiltonian, follows the instantaneous eigenstate $\ke{\psi_j(t)}$,  which derive from the initial state by continuity, when the corresponding Hamiltonian is deformed adiabatically~\cite{Messiah}. Condition for the adiabatic evolution is that  during the evolution the probability for the  transition form the eigenstate $\ke{\psi_j(t)}$ to a different one $\ke{\psi_k(t)}$ ($\forall k$) is negligible, this can be estimated as~\cite{Messiah}
\begin{eqnarray}
\sum_{k\neq j}\abs{\hbar\frac{\br{\psi_k(t)}\partial H(t)/\partial t\ke{\psi_j(t)}}{\pq{E_k(t)-E_j(t)}^2}}^2\ll1.
\end{eqnarray}
It means that larger is the difference in energy between the eigenstate state $\ke{\psi_j(t)}$ and all the other, more easily the adiabatic condition can be satisfied.

In particular if initially the system is prepared in the ground state then it will remain in the instantaneous ground state under a slow variation of some Hamiltonian parameters.
This idea can be applied to prepare the ground state of complicated Hamiltonians: One can first prepare the ground state of a sufficiently simple one which is easy to prepare. Then the Hamiltonian is adiabatically changed until approaching the final target Hamiltonian. Correspondingly the system will end up in the ground state of the final Hamiltonian.

In our case according to the result of Sec.~\ref{Strobo}, we are able to generate the dynamics corresponding to the Hamiltonian (see Eq.~\rp{Heff})
\begin{eqnarray}\label{Hefft}
H_{\rm eff}\pt{\alpha,b}&=&\beta(t)\sum_j b(t)\pq{\sigma_j^z+\alpha(t)\ \sigma_j^x}
\nn\\&&
-\beta(t)\sum_{i,j}J_{ij}\pq{\sigma_i^z\sigma_j^z +\alpha(t)\ \sigma_i^x\sigma_j^x}.
\end{eqnarray}
where $\beta(t)$ is function of $\alpha(t)$ as specified in Eq.~\rp{alphabeta}.

We want to prepare the ground state of $H_{XX}\equiv H_{\rm eff}(1,0)=-\beta\sum_{i,j}J_{ij}\pt{\sigma_i^z\sigma_j^z+\sigma_i^x\sigma_j^x}$. Hence we can first prepare the ground state of a ferromagnetic Ising Hamiltonian $H_{\rm Ising}\equiv H_{\rm eff}\pt{0,b_0}=\beta\pt{ b_0\sum_j \sigma_j^z-\sum_{i,j}J_{ij}\sigma_i^z\sigma_j^z }$ which simply corresponds to the ferromagnetic state in which all the spins are polarized along $z$. Then the ground state of $H_{XX}$ is obtained by the adiabatic variation of the parameters ${b}/{b_0}:\ 1\to 0 $ and $\alpha:\ 0\to 1$.

An example of adiabatic preparation of the ground state of the Hamiltonian $H_{\rm eff}(1,0)$ is shown in Fig~\ref{fig:adiab}. The parameters $\alpha$ and $b$ are varied according to (see the curves  Fig.~\ref{fig:adiab} (b))
\begin{eqnarray}\label{alphaht}
b(t)&=&b_0\ \ee^{-r t}
\nn\\
\alpha(t)&=&1-\ee^{-r t}.
\end{eqnarray}
Initially the parameters can be varied rapidly because the corresponding gap between ground and first excited states, is relatively large as depicted in Fig.~\ref{fig:adiab} (a). As $H_{\rm eff}$ approaches the target Hamiltonian the gap reduces and correspondingly the variation have to slow down. The final gap obtained  for $\alpha=1$ and $b=0$ and for the parameters of Fig.~\ref{fig:adiab}, is ${\rm Gap}/\hbar=21 Hz$.
The curves in Fig.~\ref{fig:adiab} (c) are obtained by numerical integration of the Schr\"odinger equation with the time dependent Hamiltonian \rp{Hefft}, and  are equal to the dashed curves in Fig.~\ref{fig:results1} (a).
The red, thin line in Fig.~\ref{fig:adiab} (c) is the fidelity between the state obtained with the adiabatic evolution and the instantaneous ground state. This curve is very close to 1 at all times indicating that the system actually follows the adiabatic ground state. The spin-spin couplings that are used in these calculation are that reported in Tab.~\ref{table:LDE}. The Hamiltonian $H_{\rm eff}(1,0)$ with these coupling strengths exhibits long range entanglement, that is strong entanglement between first and last spin. This feature is described by the blue tick curve in Fig.~\ref{fig:adiab} (c), that displays the entanglement, as measured by the concurrence between first and last spins. As expected, at large time the end spins are strongly entangled.


\end{document}